\documentstyle[12pt,preprint,aps,floats,epsf]{revtex} 
\tighten 

\begin{document} 
\preprint{ 
\noindent 
\begin{minipage}[t]{6in} 
\begin{flushright} 
KEK-TH-831 \\
hep-ph/0207184
\end{flushright} 
\begin{center} 
\end{center} 
\end{minipage} 
} 

\title{Novel constraints on $\Delta L=1$ interactions
       from neutrino masses}
\author{Francesca~Borzumati and Jae~Sik~Lee} 
\address{Theory Group, KEK, Tsukuba, Ibaraki 305-0801, Japan}   

\maketitle 

\begin{abstract} 
We reanalyze the constraints imposed on lepton-number violating
interactions by radiative contributions to neutrino masses at the one-
and two-loop levels in supersymmetric models without $R$-parity. The
interactions considered are the $\Delta L=1$ superpotential operators
$\lambda_{ijk} L_i L_j E^c_k$ and 
$\lambda^\prime_{ijk} L_i Q_j D^c_k$, and the $\Delta L=1$ soft terms
$A_{ijk} \tilde{L}_i \tilde{L}_j \tilde{E}^c_k$ and 
$A_{ijk}^\prime \tilde{L}_i \tilde{Q}_j \tilde{D}^c_k$.  The two-loop
contributions analyzed are those induced by the radiatively-generated 
mass splitting between the CP-even and CP-odd sneutrino states. It is
shown how the constraints on the couplings $\lambda_{ijk}$ and
$\lambda^\prime_{ijk}$ coming from the one-loop analysis can be
evaded.  In such a case, the two-loop contributions to neutrino masses
become important. The combined one- and two-loop analysis yields
constraints on the couplings $\lambda_{i33}$ and
$\lambda^\prime_{i33}$ that are rather difficult to escape. The
two-loop analysis yields also constraints on $A_{i33}$ and
$A^\prime_{i33}$, which are not bounded at the one-loop level.  More
freedom remains for the couplings $\lambda_{ijk}$,
$\lambda^\prime_{ijk}$, and $A_{ijk}$, $A^\prime_{ijk}$, when $j$ and
$k$ are first- or second-generation indices.
\end{abstract} 

\pacs{xxx.yy} 
\vfill  

\setlength{\parskip}{1.01ex} 

\section{Introduction}
\label{intro}
As is well known, $R_p$-violating models~\cite{RPV1,RPV2} provide a way
to generate Majorana neutrino masses without having to introduce new
fields in addition to those present in the Minimal Supersymmetric
Standard Model.  In general, however, violations of $R_p$ imply not
only violations of the lepton number $L$, but also violations of the
baryon number $B$.  This situation is dangerous as it induces a too
fast proton decay. One way to deal with this problem is to assume that
$B$ is conserved, and that $R_p$ is broken through the violation of
$L$ only. Such a choice is theoretically motivated in the context of
unified string theories~\cite{IBANEZ}. Lately, it has also received 
quite some attention in studies of collider
signatures~\cite{EFP,BKP,B-SEWS,HUITU,CHUNneutr,BGKT,BLT}.

Among the $R_p$-violating couplings that break only $L$,
$\lambda_{i33}$ and $\lambda^\prime_{i33}$ seem particularly
interesting.  Indeed, in addition to 
giving among the largest contributions
to neutrino masses~\cite{BLT}, they lead to the production of
charged~\cite{BKP} and neutral
sleptons~\cite{EFP,BKP,B-SEWS,HUITU,BLT} that may not be distinguished
from neutral and charged Higgs bosons~\cite{BLT}.  It is therefore
very important to determine how large a value for such couplings is
allowed by existing experimental results.  Direct searches of
sparticles production and of particle/sparticles decays induced by
these couplings do not constrain them very significantly (see
discussion in Ref.~\cite{BGKT}).  Indirect probes lead to constraints
that can, in general, be evaded.  This is because several other
parameters are usually involved in their extraction, whose approximate
vanishing, instead of that of the couplings $\lambda_{i33}$ and
$\lambda^\prime_{i33}$, may be responsible for the lack of any signal.

Neutrino physics, in particular, is considered one of the most severe
tests for $R_p$-violating couplings.  Hard bilinear terms from the
superpotential and soft bilinear terms are both compelled to be tiny
by the requirement that tree-level contributions to neutrino masses
are $\lesssim 1\,$eV \cite{BILINEAR}.  Similarly, it is believed that for
the one-loop contributions not to exceed the $1\,$eV mark, the
couplings $\lambda_{i33}$ and $\lambda^\prime_{i33}$ must be 
$\lesssim 10^{-4}-10^{-3}$.  Nevertheless, irrespectively of the
mechanism chosen to keep the tree-level contributions small, the
one-loop contributions can be sufficiently reduced by the requirement
that the nearly vanishing parameters are the left-right mixing terms
in the sfermion mass matrices, instead than the couplings
$\lambda_{i33}$ and $\lambda^\prime_{i33}$.  Furthermore, heavy third
generation squark masses may help suppressing the one-loop
contributions induced by the couplings $\lambda^\prime_{i33}$. All in
all, the possibility of observing charged and neutral sleptons in
incoming collider experiments, does not seem to be jeopardized by
neutrino physics, at least at the one-loop level~\cite{BLT}. It is in
this spirit that studies of such signals have been performed, for
relatively large values of the trilinear superpotential couplings
$\lambda_{i33}$ and $\lambda^\prime_{i33}$~\cite{BKP,B-SEWS,BGKT,BLT}.

There exist also two-loop contributions to neutrino masses. They are
usually ignored, since loop-suppressed with respect to the one-loop
contributions.  However, once the one-loop contributions 
are reduced down to values compatible with experimental
observations, it is not possible to neglect them anymore.  The
combinations of various parameters entering in the calculation of the
two-loop diagrams are different from those encountered in the
calculation of the one-loop diagrams. Thus, it is possible that not
all two-loop contributions are affected by the one-loop constraints
and that some of them are still rather large.  It is therefore
interesting to investigate the two-loop contributions and to establish
whether they induce additional constraints on $R_p$-violating
couplings, possibly in combination with other supersymmetric
parameters.

Some of the two-loop contributions, i.e. those proportional to the
soft trilinear $R_p$-violating couplings $A_{ijk}$ and
$A^\prime_{ijk}$, were for the first time considered in
Ref.~\cite{BFPT}. In the scenarios described there, one-loop
contributions are absent, due to symmetries forbidding the
lowest-order $R_p$-violating superpotential operators. A related
discussion can be found in Ref.~\cite{FRERE}.

In this paper, after a brief review of the one-loop contributions to
neutrino masses in Sec.~\ref{oneloop}, we analyze in detail the
two-loop contributions that are induced by the radiatively generated
splitting in the mass between CP-even and CP-odd sneutrino states; see
Sec.~\ref{twoloop}.  In particular, we give approximated formulae for
the contributions proportional to the couplings $\lambda^\prime_{i33}$
and $\lambda_{i33}$ and for those proportional to the couplings
$A^\prime_{i33}$ and $A_{i33}$.  In Sec.~\ref{numerics} we extract the
constraints that are induced on these couplings by the requirement of
neutrino masses $\lesssim 1\,$eV, making use of the combined one- and
two-loop analysis.  They turn out to be quite strong and more
difficult to evade than those obtained through the one-loop analysis
only.  Finally, we comment on the case of couplings
$\lambda^\prime_{ijk}$, $\lambda_{ijk}$, and $A^\prime_{ijk}$,
$A_{ijk}$, where $j$ and $k$ are first- or second-generation indices
and we discuss whether modifications in our results are to be expected
once the complete two-loop analysis is performed.  We conclude in
Sec.~\ref{conclusions}.

\section{One-loop analysis}
\label{oneloop}
To begin with, it is useful to review the results obtained at the one
loop. We leave aside the expression for the tree-level contributions,
which involve superpotential and scalar potential bilinear couplings
that we assume to be small at the tree level~\footnote{Bilinear terms
 can be generated radiatively at the one loop from the trilinear terms. 
 They give rise to neutrino mass terms at the two-loop 
 level. These contributions are, however, smaller than those presented 
 in the next section.}. The superpotential terms relevant for this
discussion are
\begin{equation}
W\ =\       -  \lambda^\prime_{lmn}L_{l}Q_{m}D^c_{n}
 - \frac{1}{2} \lambda_{ijk} L_i L_j E^c_k \,,
\label{trilinearsuper}
\end{equation}
whereas the $L$-violating terms in the scalar potential have the
form
\begin{equation}
 V\ =\            A^\prime_{ijk} \tilde{L}_i \tilde{Q}_j \tilde{D}^c_k
   +  \frac{1}{2} A_{ijk} \tilde{L}_i \tilde{L}_j \tilde{E}^c_k \,.
\label{trilinearsoft}
\end{equation}
The superpotential terms in Eq.~(\ref{trilinearsuper}) induce one-loop
diagrams giving rise to neutrino mass terms, with quark-squark
exchange and lepton-slepton exchange.  The largest contributions are 
due to bottom-sbottom and tau-stau exchanges.  The result of the
calculation of these diagrams is well known~\cite{ONELOOP}. The 
$b-\tilde{b}$ diagram yields
\begin{equation}
m_{\nu,i i^\prime} \ =\ \frac{3}{8\pi^2} 
    \,     \lambda^\prime_{i33} \lambda^\prime_{i^\prime 33}
    \,      m_b \, m^2_{\tilde{b},LR} 
    \, I(m_{\tilde{b}_1}^2,m_{\tilde{b}_2}^2,m_b^2) \,,         
\label{m_oneloop-b}
\end{equation}
where $m_{\tilde{b}_1}$ and $m_{\tilde{b}_2}$ are the two sbottom
eigenvalues, and $m_{\tilde{b},LR}^2$ is the left-right entry in the
$2\times 2$ sbottom mass squared matrix. The function
$I(m_{\tilde{b}_1}^2,m_{\tilde{b}_2}^2,m_b^2)$ is defined, for 
example, in Ref.~\cite{BFPT}, where also some of its limiting 
expressions are listed. It is used here in
the approximation 
$m_b/m_{\tilde{b}_1}\simeq m_b/m_{\tilde{b}_2}\simeq 0$, see 
Appendix~\ref{functions}. In the limit 
$m_{\tilde{b}_1} \to m_{\tilde{b}_2} \equiv m_{\tilde{b}}$, it 
reduces to $1/m^2_{\tilde{b}}$. Similarly, the $\tau-\tilde{\tau}$
diagram leads to:
\begin{equation}
m_{\nu,i i^\prime}\ =\ \frac{1}{8\pi^2} 
    \,     \lambda_{i33} \lambda_{i^\prime 33}
    \,     m_\tau \, m^2_{\tilde{\tau},LR} 
    \, I(m_{\tilde{\tau}_1}^2,m_{\tilde{\tau}_2}^2,m_\tau^2)\,,         
\label{m_oneloop-tau}
\end{equation}
where conventions as those for the $b-\tilde{b}$ diagram are adopted.
Notice that in both Eqs.~(\ref{m_oneloop-b})
and~(\ref{m_oneloop-tau}), no intergenerational mixing terms among
sfermions are assumed to be present.

As discussed also in Ref.~\cite{BLT}, realistically small values for
$m_{\nu,i i^\prime}$, i.e. not exceeding $1\,$eV, can be obtained 
if the following are true.
\begin{itemize}
\item[{\it (1)}]
$\lambda^\prime_{i33} \lambda^\prime_{i^\prime 33}$ and 
$\lambda_{i33} \lambda_{i^\prime 33}$ are small. Typically, 
to suppress the $b-\tilde{b}$ diagram, values as tiny as 
\begin{equation}
\left\vert
\left( \lambda^\prime_{i33}\lambda^\prime_{i^\prime 33} \right)
\left(\frac{m^2_{\tilde{b},LR}}{m^2_{\tilde{b}}}\right) 
\right\vert
 \lesssim 10^{-8} 
\label{oneloopoption1}
\end{equation}
are needed. For this estimate the two sbottom mass eigenvalues were
assumed to be of the same order of magnitude, i.e.
$m_{\tilde{b}_1}\simeq m_{\tilde{b}_2}\equiv m_{\tilde{b}}$.  Note
that, for $m^2_{\tilde{b},LR} \sim m^2_{\tilde{b}}$, the product
$(\lambda^\prime_{i33}\lambda^\prime_{i^\prime 33})$ is bound to be
$\lesssim 10^{-8}$. This constraint is eased to the value 
$\lesssim 10^{-6}$, if $m^2_{\tilde{b},LR}\sim m_b m_{\tilde{b}}$ and
$m_{\tilde{b}} \simeq 300\,$GeV. Similar considerations hold for the
$\tau-\tilde{\tau}$ diagram. A value only a factor of 5 larger than
that in the right hand side of Eq.~(\ref{oneloopoption1}) bounds from
above the combination
$(\lambda_{i33}\lambda_{i^\prime 33}) 
(m^2_{\tilde{\tau},LR}/m^2_{\tilde{\tau}})$.
\item[{\it (2)}]
$m^2_{\tilde{b},LR}$ and $m^2_{\tilde{\tau},LR}$ are small. 
For couplings $\lambda^\prime$ of ${\cal O}(1)$, and
$m_{\tilde{b}_1} \simeq m_{\tilde{b}_2} \equiv m_{\tilde{b}}$ it is
\begin{equation}
\left\vert 
 \frac{m^2_{\tilde{b},LR}}{m_{\tilde{b}}^2} 
\right\vert \lesssim 10^{-8} \,.
\label{oneloopoption2}
\end{equation}
A similar bound is obtained for 
$m^2_{\tilde{\tau},LR}/m_{\tilde{\tau}}^2$, when the couplings 
$\lambda$ are of ${\cal O}(1)$, and 
$m_{\tilde{\tau}_1}\simeq m_{\tilde{\tau}_2}\equiv m_{\tilde{\tau}}$. 
\item[{\it (3)}]
a tuning of phases in the parameters $\lambda^\prime$ and $\lambda$ 
allows a near cancellation of the two contributions.  Again, for
$m_{\tilde{b}_1} \simeq m_{\tilde{b}_2} \simeq m_{\tilde{b}}$, 
as well as 
$m_{\tilde{\tau}_1} \simeq m_{\tilde{\tau}_2}
                    \simeq m_{\tilde{\tau}}$, we obtain  
\begin{equation}
\left\vert
\left(\lambda^\prime_{i33} \lambda^\prime_{i^\prime 33} + 
      \alpha \,
      \lambda_{i33} \lambda_{i^\prime 33}
\right) \left(\frac{m^2_{\tilde{b},LR}}{m^2_{\tilde{b}}}\right)
\right\vert
 \lesssim 10^{-8}\,,
\label{oneloopoption3}
\end{equation}
where $\alpha$ is
\begin{equation} 
\alpha = 
 \frac{m_\tau}{3 m_b}
 \left(\frac{m^2_{\tilde{b}}}{m^2_{\tilde{b},LR}}\right)
 \left(\frac{m^2_{\tilde{\tau},LR}}{m^2_{\tilde{\tau}}}\right) \,.
\label{alpha}
\end{equation}
Notice that, if $m^2_{\tilde{b},LR}\sim m_b m_{\tilde{b}}$ and 
$m^2_{\tilde{\tau},LR}\sim m_\tau m_{\tilde{\tau}}$, and both 
products of couplings $\lambda$ and $\lambda^\prime$ are of 
${\cal O}(1)$, an
overall scale of the sbottom system 10 times larger than that of the
stau system is required.
\end{itemize}
Of course, all three suppression mechanisms, or two of them,
may concur to reduce the value of neutrino mass terms, therefore
alleviating the severity of constraints obtained when only one
mechanism is acting.  In the following, we shall consider 
option~{\it (3)} as the least likely among the three possibilities
listed above.  Thus, we assume that all $R_p$-violating couplings are
real, and although not necessary, we also take them to be positive.

One observation that comes out clear from this discussion, and that it
is often not appreciated enough, is that the constraints from neutrino
masses on the hard superpotential trilinear $R_p$-violating couplings,
$\lambda$ and $\lambda^\prime$, depend strongly on the details of
supersymmetry breaking. This is the obvious consequence of the fact
that the neutrino mass itself is strongly linked to supersymmetry and
supersymmetry breaking in $R_p$-violating models.  (See also
discussion in Refs.~\cite{BKP} and~\cite{NIRbook}.)  This link,
crucial at the one-loop level, will play an equally important role in
the determination of constraints for the soft trilinear
$R_p$-violating couplings, $A$ and $A^\prime$, at the two-loop level.

\section{Sneutrino mass-splitting and two-loop neutrino masses}
\label{twoloop}
There are additional one-loop diagrams contributing to neutrino
masses due to the exchange of sneutrino-neutralino,
$\tilde{\nu}-\tilde{\chi}^0$, if a mass splitting for the two physical
sneutrino states exists at the tree level~\cite{SNEUsplit0}.

As the neutral Higgs, for each generation $i$, sneutrinos have 
CP-even ($\tilde{\nu}_{i,+}$) and CP-odd components ($\tilde{\nu}_{i,-}$):
\begin{equation}
\tilde{\nu}_i \ =\  \frac{1}{\sqrt{2}} 
 \left( \tilde{\nu}_{i,+} + i \tilde{\nu}_{i,-} \right) \,,
\label{sneucurrentstates}
\end{equation}
which get equal mass from the soft mass term for $\tilde{L}_i$:
\begin{equation}
V\ =\          m_{\tilde{L}_i}^2\,\tilde{\nu}_i^\ast \tilde{\nu}_i\ =\ 
 \frac{1}{2}\, m_{\tilde{L}_i}^2\,\tilde{\nu}_{i,+} \tilde{\nu}_{i,+}  + 
 \frac{1}{2}\, m_{\tilde{L}_i}^2\,\tilde{\nu}_{i,-} \tilde{\nu}_{i,-}\,.
\label{sneutreemass}
\end{equation}
The $D$-term contributions to the sneutrino masses are considered
included in the three parameters $m_{\tilde{L}_i}^2$. For simplicity, 
we also assume these to be equal:
\begin{equation}
 m_{\tilde{L}_1}^2 = m_{\tilde{L}_2}^2 = m_{\tilde{L}_3}^2 \equiv 
 m_{\tilde{L}}^2 \,.
\label{equalslepmass}
\end{equation}
This is
not an oversimplifying assumption, since it captures the physics of
most supersymmetric models at not too large $\tan \beta$. 
Nevertheless, it does simplify significantly the following
formulas.

In general, due to the presence of bilinear terms in the
superpotential and the scalar potential, the CP-even components
$\nu_{i,+}$, in general, mix with the CP-even Higgs fields, $h$ and
$H$, and the CP-odd components $\nu_{i,-}$, mix with the CP-odd Higgs
field, $A$. Thus, a mass splitting for the CP-even and CP-odd states
is, indeed, generated at the tree level. The two one-loop diagrams
with exchange of $\tilde{\nu}_{i,+} -\tilde{\chi}^0$ and 
$\tilde{\nu}_{i,-} -\tilde{\chi}^0$, which would cancel each other if
such a splitting would not exist, give then a finite contribution to
neutrino mass terms. Such one-loop contributions are strictly related
to parameters involved also in the generation of tree-level
contributions to neutrino masses~\cite{SNEUsplitBIL-TREE}. Since we
have assumed that all bilinear $R_p$-violating terms are small at the
tree-level, we can safely neglect these contributions. (Tree-level 
mass splitting for the CP-even 
and CP-odd states are, in general, present in $R_p$-conserving models 
with right-handed neutrinos~\cite{SNEUsplitRNU-TREE}.)

\begin{figure}
\begin{center}
\epsfxsize= 14.1cm
\leavevmode
\epsfbox{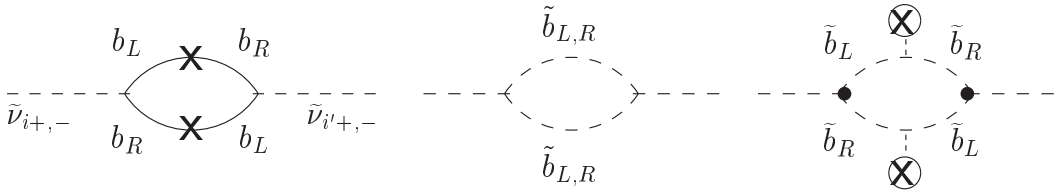}
\end{center}
\caption[f1]{\small{One-loop diagrams giving rise to a mass splitting
 between the states $\tilde{\nu}_{i,+}$ and $\tilde{\nu}_{i^\prime,-}$, 
 with $i,i^\prime = 1,2,3$. The first two diagrams are
 $\lambda^\prime$-induced, the third one, $A^\prime$-induced.}}
\label{dsneusplitting}
\end{figure}

Interactions such as those in Eqs.~(\ref{trilinearsuper})
and~(\ref{trilinearsoft}), in contrast, allow mass-splitting terms
only at the one-loop level~\cite{SNEUsplitLOOP,BLT}.  It is these
terms that are of concern for this paper and will be discussed in some
detail hereafter. Of the relevant one-loop diagrams, only those with
exchange of the $b$-quark and of $\tilde{b}$-squarks are shown in
Fig.~\ref{dsneusplitting}. There is a corresponding set of diagrams
with exchange of the $\tau$-lepton and of $\tilde{\tau}$-sleptons.
%
All these diagrams are quadratically divergent. Once the infinities
are removed through a renormalization procedure, the finite parts,
evaluated at the sneutrino scale itself, taken here to be
$m_{\tilde{L}}^2$, provide corrections to the elements of the
sneutrino mass matrix.  For states with definite CP $\sigma$ 
($\sigma = +1,-1$), these become
\begin{equation}
 m^2_{\tilde{\nu}_{ii^\prime,\sigma}} = 
 m^2_{\tilde{L}}\delta_{i i^\prime} 
 +\epsilon^2_{i i^\prime, \sigma}\,.
\label{corrsneumass}
\end{equation}
These corrections are, in general, different for states with different
$\sigma$. Indeed, the sneutrino interactions involved in the diagrams
of Fig.~\ref{dsneusplitting} are different for sneutrino states with 
different $\sigma$, as an inspection of the Lagrangian terms listed in
Appendix~\ref{interactions} shows.  Notice that no splitting is
obtained from tadpole diagrams with virtual exchange of 
$\tilde{b}$ and $\tilde{\tau}$ states.
This is because the quartic scalar interactions inducing these 
diagrams,
$\tilde{f}_L^\ast \tilde{f}_L \tilde{\nu}_i^\ast \tilde{\nu}_i$ and 
$\tilde{f}_R^\ast \tilde{f}_R \tilde{\nu}_i^\ast \tilde{\nu}_i$
($f=b,\tau$), are equal for sneutrino states with different CP.  Thus,
the finite parts of the tadpole diagrams give rise to small shifts in
the sneutrino mass terms that are identical for states with different
CP and, therefore, irrelevant for our discussion.  In the following
analysis, we neglect them altogether.

\begin{figure}
\begin{center}
\epsfxsize= 5.8cm
\leavevmode
\epsfbox{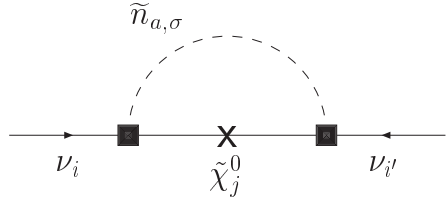}
\end{center}
\caption[f1]{\small{Contribution to the Majorana neutrino mass due to
 one-loop corrected states $\tilde{n}_{a,\sigma}$.  The couplings
 $\tilde{\chi}_{j}^0-\nu_{i}-\tilde{n}_{a,\sigma}$ indicated here by
 a square, are given explicitly in Appendix~\ref{interactions}.}}
\label{dnu1loopsneumass}
\end{figure}

Once a splitting in mass for sneutrino states with different CP is
generated, the one-loop diagram with neutralino-sneutrino exchange,
shown in Fig.~\ref{dnu1loopsneumass}, can produce nonvanishing
contributions to neutrino masses.  In this case, however, since the
sneutrino mass splitting terms are induced at the one loop, this is in
reality a diagram arising at the two-loop level. We postpone this
discussion to a later point of this paper and we proceed to the
calculation of the diagram in Fig.~\ref{dnu1loopsneumass}, where the
neutrino-sneutrino-neutralino vertices are one-loop corrected. As
Appendix~\ref{interactions} shows, this simply means that these
vertices are now weighted by the elements of the orthogonal rotation
matrices needed to obtain the sneutrino mass eigenstates from the
current eigenstates $\tilde{\nu}_{i,\sigma}$. We denote the sneutrino
mass eigenstates by the symbol $\tilde{n}_{a,\sigma}$ ($a=1,2,3$) and
the two rotation matrices by $O^\sigma$:
\begin{equation}
 \tilde{\nu}_{i,\sigma}\ =\ 
 \sum_{a} O_{ia}^{\,\sigma}\, \tilde{n}_{a,\sigma}\,.
\label{sneurotation}
\end{equation}
These two matrices are determined by the one-loop corrections
$\epsilon_{ii',\sigma}^2$.  Because of the assumption in
Eq.~(\ref{equalslepmass}), the new states $\tilde{n}_{a,\sigma}$ have
mass:
\begin{equation}
 m^2_{\tilde{n}_{a,\sigma}}  \ = \
\sum_{k, k^\prime}
   O_{ka}^{\,\sigma} m^2_{\tilde{\nu}_{kk^\prime,\sigma}} 
   O_{k^\prime a}^{\,\sigma} 
\ = \
 m^2_{\tilde{L}} + 
\sum_{k, k^\prime}
   O_{ka}^{\,\sigma} \epsilon^2_{kk^\prime, \sigma}
   O_{k^\prime a}^{\,\sigma} 
\,.
\label{sneumassrel}
\end{equation}
For the calculation of the diagram in Fig.~\ref{dnu1loopsneumass}, the
momentum dependence in the loop-corrected part of the sneutrino mass 
terms must be reinstated back explicitly.  For neutrino mass terms 
evaluated at vanishing momentum square, we obtain
\begin{equation}
 m_{\nu_{i i^\prime}} =  i \,\frac{g^2}{4} \sum_{j,a} 
 \left( \delta_{j\tilde{W}}+\delta_{j\tilde{B}} \tan^2\theta_W \right) 
 \int \frac{d^4 p}{(2\pi)^4}
 \frac{m_{\tilde{\chi}^0_j}}{p^2 -m_{\tilde{\chi}^0_j}^2} 
\left\{
 \frac{O^+_{ia} O^+_{i^\prime a}(p^2)}{p^2 -m^2_{\tilde{n}_{a,+}}(p^2)} - 
 \frac{O^-_{ia} O^-_{i^\prime a}(p^2)}{p^2 -m^2_{\tilde{n}_{a,-}}(p^2)} 
\right\} \,,
\label{neumassfirst}
\end{equation}
where $g$ is the weak coupling constant, $\theta_W$ the electroweak
mixing angle,
and the sum over $j$ extends to the two gaugino-like neutralinos. 
These are assumed to be nearly pure gauginos, so that the
neutralino diagonalization matrix reduces to the unity matrix.
An expansion of each term in the curly bracket in powers of 
the corresponding $\epsilon^2_{ii^\prime,\sigma}$,
gives
\begin{equation}
 \sum_{a}
\frac{O^\sigma_{ia} O^\sigma_{i^\prime a}(p^2)}
     {p^2 -m^2_{\tilde{n}_{a,\sigma}}(p^2)}  = 
 \frac{\delta_{ii^\prime}}{p^2 -m^2_{\tilde{L}}}  +  
 \frac{\epsilon^2_{i i^\prime,\sigma}(p^2)}
         {(p^2 -m^2_{\tilde{L}})^2} + ...  \,, 
\label{propexpansion}
\end{equation}
with the ellipsis denoting higher order terms in 
$\epsilon^2_{i i^\prime,\sigma}$. Thus, Eq.~(\ref{neumassfirst}) 
reduces to 
\begin{equation}
 m_{\nu_{i i^\prime}} =  i \,\frac{g^2}{4} \sum_{j} 
 \left( \delta_{j\tilde{W}}+\delta_{j\tilde{B}} \tan^2\theta_W \right) 
     m_{\tilde{\chi}^0_j} 
 \int \frac{d^4 p}{(2\pi)^4} 
 \frac{\delta m^2_{\tilde{\nu}_{i, i^\prime}}(p^2)}
      {(p^2 -m_{\tilde{\chi}^0_j}^2) (p^2 -m_{\tilde{L}}^2)^2}\,,
\label{genformula}
\end{equation}
where the quantities $ \delta m^2_{\tilde{\nu}_{i,i^\prime}}(p^2)$,
defined as
\begin{equation}
 \delta m^2_{\tilde{\nu}_{i,i^\prime}}(p^2)  \ \equiv \ 
\epsilon^2_{ii^\prime,+}(p^2) -\epsilon^2_{ii^\prime,-}(p^2) \,,
\label{splittingdef}
\end{equation}
are nothing but the splitting in the mass of the CP-even and CP-odd 
$i i^\prime$ sneutrino states, at the current eigenstate level.

A calculation of the diagrams in Fig.~\ref{dsneusplitting}, yields
results for these sneutrino mass splitting terms that can be
expressed, in all generality, in terms of $B$ functions (see
Appendix~\ref{analresults}). To show more explicitly their dependence
on supersymmetric parameters, we give here approximated expressions,
obtained in the limit $p^2 = 0$, separately for the $\lambda^\prime$-
and the $A^\prime$-induced terms:
\begin{eqnarray} 
 \left(\delta m^2_{\tilde{\nu}_{i,i^\prime}}(0)
 \right)\vert_{\lambda^\prime} 
&=& 
 \frac{3}{4\pi^2} \,\lambda^\prime_{i33} \lambda^\prime_{i^\prime 33}\,
 m^2_b \
 \ln\left(\frac{m^2_{\tilde{b}_2} m^2_{\tilde{b}_1}}{m_b^4}
    \right)\,,
\label{0splittlamprime}
\\
\left(\delta m^2_{\tilde{\nu}_{i,i^\prime}}(0)
 \right)\vert_{A^\prime}
&=& 
 \frac{3}{4 \pi^2} \,A^\prime_{i33} A^\prime_{i^\prime 33} \,
 \left( m^2_{\tilde{b},LR}\right)^2
   S(m^2_{\tilde{b}_1}, m^2_{\tilde{b}_2})\,.
\label{0splittAprime}
\end{eqnarray}
The first contribution has a mild logarithmic dependence on sfermion
masses; the second, has a power dependence on them. The function 
$ S(m^2_{\tilde{b}_1}, m^2_{\tilde{b}_2})$ is defined in 
Appendix~\ref{functions} and has the limiting value
$-1/(6m^4_{\tilde{b}})$ for 
$m_{\tilde{b}_1} \to m_{\tilde{b}_2}\equiv m_{\tilde{b}}$.  Similar
contributions are obtained from diagrams with $\tau-\tilde{\tau}$
exchange, induced by the couplings $\lambda_{i33}$ and $A_{i33}$. The
results for these diagrams can be read off those in
Eqs.~(\ref{0splittlamprime}) and~(\ref{0splittAprime}), after removing
the color factor $3$, and making the obvious replacements of 
couplings and masses. The overall sneutrino mass splitting is then
given by the sum of the two contributions in
Eqs.~(\ref{0splittlamprime}) and~(\ref{0splittAprime}), plus the two
contributions coming from diagrams with $\tau-\tilde{\tau}$ exchange.

It is interesting to notice that the parameters $m^2_{\tilde{b},LR}$
and $\lambda^\prime_{i33} \lambda^\prime_{i^\prime 33}$ enter in two
different contributions to the sneutrino mass splitting.  The same 
holds for the parameters $m^2_{\tilde{\tau},LR}$ and
$\lambda_{i33} \lambda_{i^\prime 33}$. Therefore, a suppression of 
the one-loop contributions to neutrino masses through option~{\it (2)}, 
among those listed after Eq.~(\ref{m_oneloop-tau}), leaves
$ \left(\delta m^2_{\tilde{\nu}_{i,i^\prime}}(p^2)
 \right)\vert_{\lambda^\prime} $ unaffected, whereas if option~{\it (1)} 
is chosen, and $m^2_{\tilde{b},LR}$ or $m^2_{\tilde{\tau},LR}$ are 
relatively large, it is 
$\left(\delta m^2_{\tilde{\nu}_{i,i^\prime}}(p^2)
 \right)\vert_{A^\prime}$ to remain unsuppressed. As for the size of 
these contributions to the sneutrino mass splitting, if we take the 
approximation 
$m_{\tilde{b}_1}    \simeq m_{\tilde{b}_2}    \equiv m_{\tilde{b}}$ 
and 
$m_{\tilde{\tau}_1} \simeq m_{\tilde{\tau}_2} \equiv m_{\tilde{\tau}}$, 
we get 
\begin{eqnarray}
\left(
\frac{\left(\delta m^2_{\tilde{\nu}_{i,i^\prime}}(0)
\right)\vert_{\lambda^\prime}}{m^2_{\tilde{L}}}
\right)
 & \simeq & 1.3 \times 10^{-3}
\left(\lambda^\prime_{i33} \lambda^\prime_{i^\prime 33}\right)
\left[\frac{\ln(m_{\tilde{b}}/m_b)}{\ln(100)}\right]
\left(\frac{100\,{\rm GeV}}{m_{\tilde{L}}}\right)^2 \,,
\label{sizeoflamprsplitt}
\\
\left(
\frac{\left(\delta m^2_{\tilde{\nu}_{i,i^\prime}}(0)
\right)\vert_{\lambda}}{m^2_{\tilde{L}}}
\right)
 & \simeq & 1.3 \times 10^{-4}
\left(\lambda_{i33} \lambda_{i^\prime 33}\right)
\left[\frac{\ln(m_{\tilde{\tau}}/m_\tau)}{\ln(56)}\right]
\left(\frac{100\,{\rm GeV}}{m_{\tilde{L}}}\right)^2 \,,
\label{sizeoflamsplitt}
\end{eqnarray}
for the $\lambda^\prime$ and $\lambda$ contributions, and 
\begin{eqnarray}
\left\vert 
 \frac{\left(\delta m^2_{\tilde{\nu}_{i,i^\prime}}(0)
 \right)\vert_{A^\prime}}{m^2_{\tilde{L}}}
\right\vert
 & \simeq & 1.3 \times 10^{-6}
\left(\frac{A^\prime_{i33} A^\prime_{i^\prime 33}}{{\rm GeV}^2}\right)
\left(\frac{m^2_{\tilde{b},LR}}{m^2_{\tilde{b}}}\right)^2
\left(\frac{100\,{\rm GeV}}{m_{\tilde{L}}}\right)^2 \,.
\label{sizeofAprsplitt}
\\
\left\vert 
 \frac{\left(\delta m^2_{\tilde{\nu}_{i,i^\prime}}(0)
 \right)\vert_{A}}{m^2_{\tilde{L}}}
\right\vert
 & \simeq & 4.2 \times 10^{-7}
\left(\frac{A_{i33} A_{i^\prime 33}}{{\rm GeV}^2}\right)
\left(\frac{m^2_{\tilde{\tau},LR}}{m^2_{\tilde{\tau}}}\right)^2
\left(\frac{100\,{\rm GeV}}{m_{\tilde{L}}}\right)^2 \,,
\label{sizeofAsplitt}
\end{eqnarray}
for the $A^\prime$ and $A$ contributions.

As we shall see, similar considerations hold for the neutrino mass
terms obtained from Eq.~(\ref{genformula}), once the expression for
the sneutrino mass splitting is substituted in, and the integral is
performed.  The exact analytic expression for $m_{\nu_{i i^\prime}}$
is cumbersome. We show here the expression obtained when 
$\delta m^2_{\tilde{\nu}_{i,i^\prime}}(p^2)$ is approximated by 
$\delta m^2_{\tilde{\nu}_{i,i^\prime}}(0)$. The results obtained
through this approximation are certainly not valid for evaluations in
which factors of ${\cal O}(1)$ are important. They are, however, good
enough for order of magnitude estimates. We get
\begin{equation} 
m_{\nu_{i i^\prime}} \simeq  \frac{g^2}{64 \pi^2} \sum_j 
\left( \delta_{j\tilde{W}}+\delta_{j\tilde{B}}\tan^2\theta_W \right)
     m_{\tilde{L}} 
\left(\!\frac{\delta m^2_{\tilde{\nu}_{i, i^\prime}}(0)}
           {m_{\tilde{L}}^2}\!
\right)
 J\left(m^2_{\tilde{\chi}^0_j},m^2_{\tilde{L}}\right)\,,  
\label{approxmass}
\end{equation}
where $\delta m^2_{\tilde{\nu}_{i, i^\prime}}(0)$ is given by the 
sum of 
$\left(\delta m^2_{\tilde{\nu}_{i,i^\prime}}(0) \right)
\vert_{\lambda^\prime} $ and 
$\left(\delta m^2_{\tilde{\nu}_{i,i^\prime}}(0)
 \right)\vert_{A^\prime}$ in Eqs.~(\ref{0splittlamprime}) 
and~(\ref{0splittAprime}). The dimensionless function 
$ J(m^2_{\tilde{\chi}^0_j},m^2_{\tilde{L}})$, defined explicitly in 
Appendix~\ref{functions}, is positive definite and depends only on 
the ratio $y\equiv m^2_{\tilde{\chi}^0_j}/m^2_{\tilde{L}}$.
Note that the critical proportionality to the neutralino mass,
explicit in Eq.~(\ref{genformula}), is now hidden by the fact that a
factor $m_{\tilde{\chi}^0_j}/m_{\tilde{L}}$ has been included in the
function $ J$. This is indeed the product of the ratio
$m_{\tilde{\chi}^0_j}/m_{\tilde{L}}$ and the actual loop function 
$I^\prime(m^2_1,m^2_2)$, also given in Appendix~\ref{functions}, and 
also depending only on the ratio $y$.  The
reason for this replacement is that the $J$ function is very slowly
decreasing for $y \gtrsim 4.5$, where it reaches the maximum value
$\simeq 0.57$. It is also very slowly decreasing from this point down
to $y=1$, where it gets the value $1/2$. It has still the value 
$\sim 1/4$ at $y=0.1$, and it drops to $0$ as $\sqrt{y}$, for 
$y \to 0$.

\begin{figure}
\begin{center}
\epsfxsize= 6.7cm
\leavevmode
\epsfbox{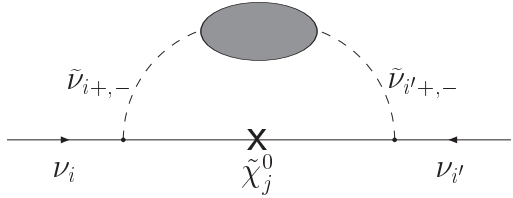}
\end{center}
\caption[f1]{\small{Two-loop contribution to Majorana neutrino mass
 terms. The grey oval indicates any of the one-loop diagrams in
 Fig.~\ref{dsneusplitting} generating a mass splitting for the
 sneutrino states $\tilde{\nu}_{i,+}$ and $\tilde{\nu}_{i^\prime,-}$.}}
\label{dnu2loopmass}
\end{figure}

As preannounced, this contribution to neutrino mass is split in two 
terms. The first, as the one-loop contribution, depends on the 
product $\lambda^\prime_{i33} \lambda^\prime_{i^\prime 33}$
($\lambda_{i33} \lambda_{i^\prime 33}$) but not on 
$m^2_{\tilde{b},LR}$ ($m^2_{\tilde{\tau},LR}$). Because of this, and 
because of the milder logarithmic dependence on scalar masses, it is
less sensitive to the details of supersymmetry breaking than the
one-loop neutrino-mass contribution. The second, on the contrary, is
very sensitive to these details. It is directly proportional to
$A^\prime_{i33} A^\prime_{i^\prime 33}$ 
($A_{i33} A_{i^\prime 33}$), which do not have any influence on 
neutrino physics at the one-loop level.

The contribution to neutrino masses just calculated was treated as
arising from one-loop diagrams with one-loop corrected vertices.
That is to say, it is in reality a contribution originating at the
two-loop level. The corresponding diagram is shown explicitly in
Fig.~\ref{dnu2loopmass}, where the grey oval is any of the one-loop
diagrams in Fig.~\ref{dsneusplitting}.  It is straightforward to see
that the procedure followed here and the direct calculation of the
two-loop diagram are completely equivalent and lead to the same
Eq.~(\ref{genformula}).

Clearly, this equation does not exhaust all possible two-loop
contributions to neutrino masses, neither all those proportional to
the product of two $\lambda^\prime$ (or $\lambda$) couplings or of
two $A^\prime$ (or $A$) couplings.  Equation~(\ref{genformula})
encapsulates only the two-loop contributions that are induced by 
the one-loop generated splitting in the mass of CP-even and CP-odd 
sneutrino states~\footnote{There exist also other diagrams, not 
 belonging to the class of diagrams considered here, that depend on
 products of trilinear couplings
 $\lambda^\prime_{i33} \lambda^\prime_{i^\prime 33}$ 
 ($\lambda_{i33} \lambda_{i^\prime 33}$) but not on 
 $m^2_{\tilde{b},LR}$ ($m^2_{\tilde{\tau},LR}$). See, for example,
 the one-particle reducible diagrams that are obtained from two
 juxtaposed one-loop neutrino-neutralino transitions. An estimate
 of these diagrams shows that they give contributions to neutrino
 masses that are not larger than those considered here, at least in 
 generic regions of the supersymmetric parameter space. We thank
 E.J. Chun for bringing this to our attention.}.
That is to say, the vertex corrections depicted by a
black box in Fig.~\ref{dnu1loopsneumass} are only due to the
correction to the mass of the different sneutrino states.
%
\begin{figure}
\begin{center}
\epsfxsize= 5.8cm
\leavevmode
\epsfbox{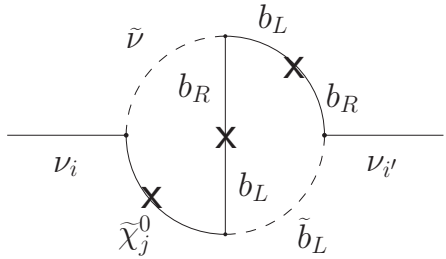}
\end{center}
\caption[f1]{\small{Another two-loop contribution to the Majorana
 neutrino mass.}}
\label{dnu2cutlemon}
\end{figure}
Genuine vertex corrections for the interaction
neutrino-sneutrino-neutralino, also due to $R_p$-violating couplings,
would give rise to two-loop diagrams with a different topology. An
example is explicitly shown in Fig.~\ref{dnu2cutlemon}. The dependence
on supersymmetric couplings in this diagram is as that of the diagram
in Fig.~\ref{dnu2loopmass}, when the grey oval stands for the first
diagram in Fig.~\ref{dsneusplitting}.  Such a diagram should not spoil
our estimate of the constraints that neutrino masses induce on
$R_p$-violating couplings. Other diagrams enter in the complete
two-loop analysis, with the same topology as the diagram of
Fig.~\ref{dnu2cutlemon}. A complete calculation should, therefore, be 
performed.

\section{Numerical Analysis}
\label{numerics}
We are now in a position to discuss the numerical implications of our
analysis.  Equation~(\ref{approxmass}) clearly bears what we
claimed earlier,
i.e.  that the two-loop contributions to neutrino masses induced by
the one-loop generated mass splitting between CP-even and CP-odd
sneutrino states are quite large. This remains so, even after the
one-loop contributions have been reduced as to avoid conflicts with
experimental results.

First of all, the requirement that the neutrino mass terms do not
exceed the $1\,$eV mark induces a constraint on the fractional
splitting in the mass of CP-even and CP-odd sneutrino states that is
independent from the $R_p$-violating couplings that have actually
induced it.  Taking the value $y=1$ for our numerical estimate, we
obtain
\begin{equation}
\left(\!\frac{\delta m^2_{\tilde{\nu}_{i, i^\prime}}(0)}
           {m_{\tilde{L}}^2}\!
\right) \ \lesssim \ 10^{-8} 
\left(\frac{100\,{\rm GeV}}{m_{\tilde{L}}}\right) \,.
\label{sneusplconstr}
\end{equation}
This constraint is rather difficult to evade. For 
$m_{\tilde{L}} \simeq 100\,$GeV, a fractional splitting of a
few$\,$GeV$^2$ would require $J(y)\sim\sqrt{y}\sim 10^{-4}$, resulting
in $10\,$MeV weak-gaugino masses, which are already excluded by
present experiments.

In turn, constraints on $R_p$-violating couplings can also be
derived, namely,
\begin{eqnarray}
\left(\lambda^\prime_{i33} \lambda^\prime_{i^\prime 33}\right)
\left[\frac{\ln(m_{\tilde{b}}/m_b)}{\ln(100)}\right]
\left(\frac{100\,{\rm GeV}}{m_{\tilde{L}}}\right) 
 &\lesssim& 10^{-5} \,,
\label{lampr2loopconstr}
\\
\left(\lambda_{i33} \lambda_{i^\prime 33}\right)
\left[\frac{\ln(m_{\tilde{\tau}}/m_\tau)}{\ln(56)}\right]
\left(\frac{100\,{\rm GeV}}{m_{\tilde{L}}}\right) 
 &\lesssim& 10^{-4} \,,
\label{lam2loopconstr}
\end{eqnarray}
for the $\lambda^\prime$- and $\lambda$-couplings, and
\begin{eqnarray}
\left(A^\prime_{i33} A^\prime_{i^\prime 33}\right)
\left(\frac{m^2_{\tilde{b},LR}}{m^2_{\tilde{b}}}\right)^2
\left(\frac{100\,{\rm GeV}}{m_{\tilde{L}}}\right)
 & \lesssim & 10^{-2} {\rm GeV}^2 \,,
\label{Apr2loopconstr}
\\
\left(A_{i33} A_{i^\prime 33}\right)
\left(\frac{m^2_{\tilde{\tau},LR}}{m^2_{\tilde{\tau}}}\right)^2
\left(\frac{100\,{\rm GeV}}{m_{\tilde{L}}}\right)
 & \lesssim & 10^{-1} {\rm GeV}^2 \,,
\label{A2loopconstr}
\end{eqnarray}
for the $A^\prime$- and $A$-couplings. Both sets of constraints depend
only linearly on the sneutrino scale. Those on the products
$\lambda^\prime_{i33} \lambda^\prime_{i^\prime 33}$ and 
$\lambda_{i33} \lambda_{i^\prime 33}$ have also a rather weak
dependence on the sbottom and on the charged slepton scale,
respectively.

If option~{\it (2)} is chosen to suppress the one-loop contribution to
the neutrino mass, i.e. if the left-right mixing terms in the sfermion
mass matrices are very small, then Eq.~(\ref{lampr2loopconstr}) bounds
the products $\lambda^\prime_{i33}\lambda^\prime_{i^\prime 33}$ to be
$\lesssim 10^{-5}$ for $m_{\tilde{b}}\sim 300\,$GeV and 
$m_{\tilde{L}}\sim 100\,$GeV. The products 
$A^\prime_{i33} A^\prime_{i^\prime 33}$ remain unbounded.  On the
contrary, if the products of couplings 
$\lambda^\prime_{i33} \lambda^\prime_{i^\prime 33}$ are constrained by
the one-loop contributions to the neutrino mass terms, then only the
products $A^\prime_{i33} A^\prime_{i^\prime 33}$ are bounded by the
two-loop analysis. They can be at most a few hundred$\,$GeV$^2$, if
$m^2_{\tilde{b},LR} \sim m_b m_{\tilde{b}}$, 
$m_{\tilde{b}} \sim 300\,$GeV, and $m_{\tilde{L}} \sim 100\,$GeV. For
$m^2_{\tilde{b},LR} \sim m^2_{\tilde{b}}$ and still 
$m_{\tilde{L}} \sim 100\,$GeV, the value of these products is reduced
to $\sim 10^{-2}$ GeV$^2$.

Similar constraints hold for the product of
couplings $\lambda_{i33} \lambda_{i^\prime 33}$ and 
$A_{i33} A_{i^\prime 33}$.

All the above discussion has been devoted to $R_p$-violating trilinear
couplings with at least two third-generation indices.  For couplings
$\lambda^\prime_{i jk}$ and $A^\prime_{i jk}$ with $j=k \neq 3$, the
analysis follows the same pattern as that for the case with $j=k=3$.
The $b$-quark and the $\tilde{b}$-squarks must be replaced by the $d$-
or $s$-quark and the $\tilde{d}$- or the $\tilde{s}$-squarks
everywhere in the above equations.  The constraints corresponding to
those in Eq.~(\ref{lampr2loopconstr}) on $\lambda^\prime$-products
with $j=k=2$ are about 3 or 4 order of magnitude weaker than those for
$\lambda^\prime_{i33}\lambda^\prime_{i^\prime 33}$, depending on the
particular value used for the mass of the $s$-quark. They are
completely lost for $j=k=1$. 
Therefore, if option~{\it (2)} is chosen to reduce the one-loop
contributions to neutrino masses, the products
$A^\prime_{i22} A^\prime_{i^\prime 22}$ and
$A^\prime_{i11} A^\prime_{i^\prime 11}$ as well as 
$\lambda^\prime_{i11}\lambda^\prime_{i^\prime 11}$ remain 
substantially unsuppressed, whereas the bound   
$\lambda^\prime_{i22}\lambda^\prime_{i^\prime 22} \lesssim 10^{-2}$ 
is obtained.  In contrast, if option~{\it (1)} is selected, the
severity of both, the one-loop constraints on the
$\lambda^\prime$-products and the two-loop ones on the
$A^\prime$-products, depends on the mechanism of supersymmetry
breaking. More precisely, it depends on whether the left-right mixing
terms $m_{\tilde{f},LR}^2$ are proportional or not to the
corresponding fermion mass $m_f$. In the former case, for a squark
mass of $300\,$GeV, it is
$\lambda^\prime_{i22} \lambda^\prime_{i^\prime 22}
                             \lesssim 10^{-3}-10^{-2}$, whereas 
$\lambda^\prime_{i11} \lambda^\prime_{i^\prime 11}\sim {\cal O}(1)$ 
are still allowed.  The two-loop constraints in
Eq.~(\ref{A2loopconstr}) on the $A^\prime$-products would still easily
allow couplings of type $A^\prime$ as large as
$100\,$GeV or $\sim 1\,$TeV for $j=k=2$, and even larger for $j=k=1$. 
For $m_{\tilde{f},LR}^2$ independent from $m_f$, the results for the
$A^\prime$ products are the same as in the case of 
$A^\prime_{i33} A^\prime_{i^\prime 33}$, the constraints on the
$\lambda^\prime$ products are:
$\lambda^\prime_{i22}\lambda^\prime_{i^\prime 22}
                              \lesssim 10^{-7}-10^{-6}$ and 
$\lambda^\prime_{i11}\lambda^\prime_{i^\prime 11}
                              \lesssim 10^{-5}-10^{-4}$.
As already mentioned, the uncertainties in the upper bounds discussed
here is due to the uncertainty in the $s$- and $d$-quark masses.

Constraints on the products of $\lambda$ and $A$ couplings with 
indices $j=k=2$ or $j=k=1$ can be be obtained in a similar way.

\begin{table}
\begin{tabular}{|c||c|c||c|c|c|c|}
&
\multicolumn{2}{c||}{$m_{\tilde{f},LR}^2\lesssim 10^{-8}\,
  \left(\displaystyle{\frac{m_b,m_\tau}{m_f}}\right)\, m^2_{\tilde f}$}&
\multicolumn{2}{c|}{$m_{\tilde{f},LR}^2=m_f\, m_{\tilde f}$}&
\multicolumn{2}{c|}{$m_{\tilde{f},LR}^2=m^2_{\tilde f}$ } 
\\
\hline
& 
1-loop & 
2-loop & 
1-loop & 
2-loop & 
1-loop & 
2-loop  
\\
\cline{2-7}
&&&&&&\\
$\lambda^\prime_{i33}\ (f\!=\!b)$ & 
-- & 
$10^{-2}$ & 
$10^{-3}$ & 
$10^{-2}$ & 
$10^{-4}$ & 
$10^{-2}$  
\\
$\lambda^\prime_{i22}\ (f\!=\!s)$ & 
-- & 
$10^{-1}$ & 
$10^{-2}$ & 
$10^{-1}$ & 
$10^{-3}$ & 
$10^{-1}$  
\\
$\lambda^\prime_{i11}\ (f\!=\!d)$ & 
-- & 
-- & 
$\sim 1$  & 
--        & 
$10^{-3}$ & 
--         
\\
\hline
$\lambda_{i33}\ (f\!=\!\tau)$ & 
-- & 
$10^{-2}$ & 
$10^{-3}$ & 
$10^{-2}$ & 
$10^{-4}$ & 
$10^{-2}$  
\\
$\lambda_{i22}\ (f\!=\!\mu)$ & 
-- & 
$10^{-1}$ & 
$10^{-2}$ & 
$10^{-1}$ & 
$10^{-3}$ & 
$10^{-1}$  
\\
$\lambda_{i11}\ (f\!=\!e)$ & 
-- & 
-- & 
-- & 
--        & 
$10^{-2}$ & 
--         
\\
\end{tabular}
\caption[]{\small{One- and two-loop upper limits on
 $\lambda^\prime_{ijj}$ and $\lambda_{ijj}$ ($i,j=1,2,3$) induced by
 the requirement that the neutrino mass terms do not exceed $1\,$eV.
 The numerical values used for quark and lepton masses are: 
 $m_b=3\,$GeV, $m_b/m_s=40$, $m_s/m_d=20$, and $m_\tau=1.8\,$GeV, 
 $m_\tau/m_\mu=20$, $m_\mu/m_e=200$. Squark masses are all 
 fixed at $300\,$GeV, slepton and neutralino masses at $100\,$GeV.  
 Different choices are made for $m^2_{\tilde{f},LR}$, from the very
 small to the rather large, see text.  A horizontal bar indicates that
 the corresponding $\lambda^\prime_{ijj}$ or $\lambda_{ijj}$ are
 allowed to be even larger than 1.}} 
\label{Table1}
\end{table}   
%
%
A summary of the upper bounds for each of the $R_p$-violating
couplings discussed so far is given in Table~\ref{Table1}
and~\ref{Table2}. In particular, Table~\ref{Table1} lists the upper
limits for the couplings $\lambda^\prime_{ijj}$ and
$\lambda_{ijj}$, Table~\ref{Table2}, those for the couplings 
$A^\prime_{ijj}$ and $A_{ijj}$.  For the numerical calculation, the
values $m_b=3\,$GeV, $m_b/m_s=40$, $m_s/m_d=20$ were used for the
quark masses, $m_\tau=1.8\,$GeV, $m_\tau/m_\mu=20$, and
$m_\mu/m_e=200$, for the lepton masses.  The squark masses were fixed
as:
$m_{{\tilde d}_{1,2}}=m_{{\tilde s}_{1,2}}=m_{{\tilde b}_{1,2}}=
       m_{\tilde q}=300\,$GeV. Finally, 
$m_{{\tilde e}_{1,2}}=m_{\tilde{\mu}_{1,2}}=m_{\tilde{\tau}_{1,2}}=
       m_{\tilde \ell}=100\,$GeV, and 
$m_{\tilde L}=m_{\tilde W}=100\,$GeV were chosen for the sleptons and 
the neutralino masses.  In both tables the constraints are given for
different possible values of $m^2_{\tilde{f},LR}$:
$m_{\tilde{f},LR}^2\lesssim 10^{-8}\,(m_b/m_f)\,m^2_{\tilde f}$, which
sufficiently suppresses the one-loop contribution to neutrino masses;
$m_{\tilde{f},LR}^2=m_f\, m_{\tilde f}$; and 
$m_{\tilde{f},LR}^2=m^2_{\tilde f}$.  Note that the 2-loop 
constraints on $\lambda$ and $\lambda^\prime$ couplings do not 
depend on $m^2_{\tilde{f},LR}$, see Eq.~(\ref{0splittlamprime}), and 
therefore are important only when option~{\it (2)} is chosen to 
suppress the one-loop contribution to neutrino mass, i.e. for
$m_{\tilde{f},LR}^2\lesssim 10^{-8} m_{\tilde{f}}^2$.
The constraints on $A$ and $A^\prime$ couplings do not depend on
$m_f$, but they depend on $m_{\tilde{f},LR}^2$. For
$m_{\tilde{f},LR}^2=m^2_{\tilde f}$, they are completely
flavor independent, see Eq.~(\ref{0splittAprime}). A horizontal bar
in both tables indicates that the corresponding coupling is completely
unconstrained. That is to say, even values larger than 1, in the case
of $\lambda^\prime_{ijj}$ or $\lambda_{ijj}$, and larger than
$1\,$TeV, in the case of $A^\prime_{ijj}$ or $A_{ijj}$ are not in
conflict with the required smallness of neutrino masses.
%
%
\begin{table}
\begin{tabular}{|c||c||c|c|}
  &
$m_{\tilde{f},LR}^2\lesssim 10^{-8}\,
 \left(\displaystyle{\frac{m_b,m_\tau}{m_f}}\right)\, m^2_{\tilde f}$ & 
$m_{\tilde{f},LR}^2=m_f\, m_{\tilde f}$ &
$m_{\tilde{f},LR}^2=m^2_{\tilde f}$ 
\\
 \hline
$A^\prime_{i33}\ (f\!=\!b)$ & 
{--}                        &
{$10\,$GeV}                 &
{$0.1\,$GeV}    
\\
$A^\prime_{i22}\ (f\!=\!s)$  & 
{--}                         &
{$\sim\! 1\,$TeV}            &
{$0.1\,$GeV}       
\\
$A^\prime_{i11}\ (f\!=\!d)$  & 
{--}                         &
{--}                         &
{$0.1\,$GeV} 
\\
\hline
$A_{i33}\ (f\!=\!\tau)$      & 
{--}                         &
{$10\,$GeV}                  &
{$0.1\,$GeV}    
\\
$A_{i22}\ (f\!=\!\mu)$       & 
{--}                         &
{$100\,$GeV}                 &
{$0.1\,$GeV}       
\\
$A_{i11}\ (f\!=\!e)$         & 
{--}                         &
{--}                         &
{$0.1\,$GeV} 
\\
\end{tabular}
\caption[]{\small{Same as in Table~\ref{Table1}, for the couplings 
 $A^\prime_{ijj}$ and  $A_{ijj}$ with $i,j=1-3$. These constraints
 are obtained at the two-loop level.  A horizontal bar indicates that
 the corresponding $A^\prime_{ijj}$ or $A_{ijj}$ are allowed to be
 even larger than $1\,$TeV.}}
\label{Table2}
\end{table}   
%

Note that an overall increase by a factor of $3$ in all sfermion
masses, i.e. $m_{\tilde{q}} \sim 1\,$TeV and 
$m_{\tilde{\ell}}=m_{\tilde{L}}=300\,$GeV, would affect the constraints
on the products of $R_p$-violating couplings, but it would not change 
the upper bounds on the $\lambda^\prime$ and $A^\prime$ couplings. It 
can, however, weaken those for $\lambda_{i22}$, $A_{i33}$, and 
$A_{i22}$ by one order of magnitude, when 
$m_{\tilde{f},LR}^2=m_f\, m_{\tilde f}$. 

The situation is more complicated for couplings in which only one
index $j$ or $k$ is $3$, while the other is not. The constraints on
the different products of couplings cannot be simply gleaned from the
results presented here and require an independent calculation. Not all
diagrams in Fig.~\ref{dsneusplitting}, for example, contribute to the
sneutrino mass splitting and the interaction terms are more involved
than those presented in Appendix~\ref{interactions}.  Such a
calculation goes beyond the scope of this paper and calls for an
independent analysis. We can only guess that the constraints in these
cases are less severe than in the case $j=k=3$, but probably more
bounding than those obtained in the cases $j=k=2$ and $j=k=1$.

We would like to stress again that the constraints just derived 
are up to coefficients of ${\cal O}(1)$. The use of 
Eq.~(\ref{genformula}), instead of Eq.~(\ref{approxmass}), gives
rise to small imaginary parts for the neutrino mass terms. They
are due to the analytic cuts that the loop functions for the 
diagrams in Fig.~\ref{dsneusplitting} have at large $p^2$. 
These aspects of the exact calculation, although interesting, 
are inconsequential for our discussion. We have explicitly 
checked that the order of magnitude of our estimates is, indeed, 
quite reliable.

\section{Conclusions}
\label{conclusions}
We conclude this paper with the following observations.  The smallness
of neutrino mass is, indeed, one of the most powerful constraints on
some of the $R_p$-violating couplings involving third-generation
indices. Contrary to what may be naively thought, the two-loop
contributions to neutrino masses are very important to constrain such
couplings.  Barring cancellations in which the phases of these
couplings play an important role, the complete one- and two-loop
analysis indicates that it is rather unlikely that the couplings
$\lambda^\prime_{i33}$ and $\lambda_{i33}$ are above the percent
level.  Such a constraint, obtained from two-loop contributions to
neutrino masses, is irrespective of the value of the left-right mixing
terms in the sbottom and stau mass matrices.  It can only become more
severe if these mixing terms are nonnegligible and the one-loop
contributions to neutrino masses are unsuppressed.  On the contrary,
the soft trilinear parameters $A^\prime_{i33}$ and $A_{i33}$, which
enter only in two-loop contributions to neutrino masses, are rather
weakly constrained.  The most stringent upper bound, obtained for very
large left-right mixing terms, is only $0.1\,$GeV.  They are
completely unbounded from above for small left-right mixing terms.

Some of the consequences that the constraints derived here have 
for collider physics are discussed in Ref.~\cite{BLT}.

\vspace*{0.5truecm}
\noindent 
{\bf Acknowledgements}  
The authors acknowledge the financial support of the 
Japanese Society for Promotion of Science.

\newpage
\appendix
%
\section{Interaction terms}
\label{interactions}
In this Appendix are listed the interaction Lagrangian terms needed
for the calculations presented in the text.  In the limit in which the
two lightest neutralinos are pure gauginos, the
neutralino-sneutrino-neutrino interactions terms are
\begin{eqnarray}
 {\cal L}_{\rm tree} & \supset & 
- \frac{g}{\sqrt{2}} 
 \sum_{i,j}
  \left(\delta_{j\tilde{W}}-\delta_{j\tilde{B}} \tan \theta_W \right) 
\left\{
  \overline{(\tilde{\chi}_{j}^0)} P_L \nu_{i}\,\tilde{\nu}_{i}^\ast + 
  \overline{\nu_{i}} P_R\tilde{\chi}_{j}^0  \,\tilde{\nu}_{i}
\right\}
\nonumber \\[1.001ex]
& = & 
- \, 
  \frac{g}{2} \, 
 \sum_{i,j} 
  \left(\delta_{j\tilde{W}}-\delta_{j\tilde{B}} \tan \theta_W \right) \,
\left\{
 \overline{(\tilde{\chi}_{j}^0)}         \nu_{i}  \tilde{\nu}_{i,+} +i 
 \overline{(\tilde{\chi}_{j}^0)}\gamma_5 \nu_{i} \tilde{\nu}_{i,-}       
\right\}\,,
\label{gaugeintertree}
\end{eqnarray}
where the fact that both neutrinos and neutralinos are Majorana
particles was used.  Once one-loop corrections to the sneutrino sector
are included, the neutralino-sneutrino-neutrino interactions terms can
be expressed in terms of the mass eigenstates $\tilde{n}_{a\sigma}$
($\sigma =+1,-1$), defined in Eq.~(\ref{sneurotation}), as
\begin{equation}
 {\cal L}_{{\rm 1-loop}\vert\tilde{\nu}} \ \supset \ 
- \frac{g}{2} 
 \sum_{i,j,a}
  \left(\delta_{j\tilde{W}}-\delta_{j\tilde{B}} \tan \theta_W \right) 
\left\{ O^+_{ia}
\overline{(\tilde{\chi}_{j}^0)} \nu_{i}
\tilde{n}_{a,+}                          + i 
        O^-_{ia}
\overline{(\tilde{\chi}_{j}^0)} \gamma_5 \nu_{i}
\tilde{n}_{a,-}                      
\right\}\,.
\label{gaugeinter1loop}
\end{equation}
At the tree level, the current eigenstates $\tilde{\nu}_{i\sigma}$ and
the mass eigenstates $\tilde{n}_{a\sigma}$ coincide.

The $R_p$-violating terms obtained from the superpotential in
Eq.~(\ref{trilinearsuper}) are
\begin{eqnarray}
 {\cal L}_{\rm tree} & \supset &  
-\sum_{i} \left\{
 \lambda_{i33}^\prime   \,
 \left(\overline{b_R}   \, b_L    \right)\tilde{\nu}_{i}        
+ \lambda_{i33}         \,
 \left(\overline{\tau_R}\, \tau_L \right)\tilde{\nu}_{i}
          \right\}  + {\rm H.c.}
\nonumber \\[1.001ex]
& = & 
-\sum_{i} \left\{
 \frac{\lambda_{i33}^\prime}{\sqrt{2}}
 \left[    \bar{b} b               \,\tilde{\nu}_{i,+} 
        -i\,\bar{b} \gamma_5 b      \,\tilde{\nu}_{i,-} \right]
+\frac{\lambda_{i33}}{\sqrt{2}}
 \left[    \bar{\tau} \tau         \,\tilde{\nu}_{i,+} 
        -i\,\bar{\tau} \gamma_5\tau \,\tilde{\nu}_{i,-} \right]
          \right\} \,.
\label{snuferferintertree}
\end{eqnarray}
Note that both sets of couplings $\lambda^\prime_{i33}$ and
$\lambda_{i33}$ are taken to be real. In other words, of the three
options listed in the text after Eq.~(\ref{m_oneloop-tau}), option 
{\it (3)} is not considered likely to be the one suppressing the
one-loop contributions to neutrino masses.

The sneutrino-sbottom-sbottom interaction terms and the 
sneutrino-stau-stau ones, respectively proportional 
to the $\lambda^\prime$ and $\lambda$ couplings, and coming  
from $F$-terms, are
\begin{eqnarray} 
 {\cal L}_{\rm tree} & \supset &  
- \sum_{i}            \left\{
 m_b \lambda_{i33}^\prime 
 \left( \tilde{b}^\ast_R \tilde{b}_R +\tilde{b}^\ast_L \tilde{b}_L 
 \right)\tilde{\nu}_i 
+ m_\tau\lambda_{i33}     
 \left( \tilde{\tau}^\ast_R \tilde{\tau}_R + 
        \tilde{\tau}^\ast_L \tilde{\tau}_L 
 \right)\tilde{\nu}_i \right\}   + {\rm H.c.} 
\nonumber \\
& = & 
- {\sqrt{2}} \sum_{i,j}
\left\{
     m_b \lambda_{i33}^\prime
 \left( \tilde{b}^\ast_j \tilde{b}_j \right)\tilde{\nu}_{i,+} 
+ m_\tau \lambda_{i33} 
 \left( \tilde{\tau}^\ast_j \tilde{\tau}_j \right)\tilde{\nu}_{i,+} 
\right\} \,.
\label{Fterminter}
\end{eqnarray}
Again, the assumption of real $\lambda_{i33}^\prime$ and
$\lambda_{i33}$ couplings was made.  Notice that no interaction terms
for the CP-odd states is generated by $F$-term contributions to the
Lagrangian.

Finally, contributions to the sneutrino-sbottom-sbottom 
interaction terms and to the sneutrino-stau-stau ones, 
come also from the scalar potential in 
Eq.~(\ref{trilinearsoft}). They are
\begin{equation}
 {\cal L}_{\rm tree} \ \supset \  
 - \sum_{i} \left\{
   A^\prime_{i33} \tilde{b}_L^\ast \tilde{b}_R \tilde{\nu}_i +
   A_{i33} 
   \tilde{\tau}_L^\ast \tilde{\tau}_R \tilde{\nu}_i
            \right\} + {\rm H.c.}\,.
\label{scalpotinter0}
\end{equation}
Once the $2\times 2$ sbottom mass squared matrix is diagonalized, the  
first term becomes
{\small
\begin{equation}
 {\cal L}_{\rm tree} = 
 - \sum_{i} \frac{A^\prime_{i33}}{\sqrt{2}} 
\left\{ \left[
 \sin 2 \theta_{\tilde{b}}  \left(
 \tilde{b}_1^\ast \tilde{b}_1 - \tilde{b}_2^\ast \tilde{b}_2
                  \right) + 
 \cos 2 \theta_{\tilde{b}}  \left(
 \tilde{b}_1^\ast \tilde{b}_2 + \tilde{b}_2^\ast \tilde{b}_1
                  \right)
        \right] \tilde{\nu}_{i,+} + 
      i \left[
 \tilde{b}_2^\ast \tilde{b}_1 - \tilde{b}_1^\ast \tilde{b}_2 
        \right] \tilde{\nu}_{i,-} 
\right\}\,,   
\label{scalpotinter}
\end{equation}
}
where $\theta_{\tilde{b}}$ is the diagonalization angle of the
$2\times 2$ sbottom mass squared matrix. A similar expression is
obtained for the sneutrino-stau-stau interaction term in
Eq.~(\ref{scalpotinter0}).

Note that, for the purpose of obtaining neutrino mass terms induced by
the splitting in mass for the CP-even and CP-odd sneutrino states, it
is sufficient to consider the effect of sneutrino mass corrections on
the neutralino-sneutrino-neutrino interaction terms only.  The
inclusion of such corrections to the other interaction terms in
Eqs.~(\ref{snuferferintertree}),~(\ref{Fterminter}),
and~(\ref{scalpotinter}) is irrelevant for the calculation of 
neutrino mass terms induced by a nonvanishing splitting in the mass 
of CP-even and CP-odd sneutrino states.

\section{Sneutrino mass-splitting terms}
\label{analresults}
We list here the formulae obtained for the sneutrino mass 
splitting from one-loop diagrams:
\begin{eqnarray} 
(\delta m_{\tilde{\nu}_{i i^\prime}}^2(p^2))^{(1)}\vert_{A^\prime}
&=& 
 - \frac{3}{4\pi^2} A^\prime_{i33} A^\prime_{i^\prime 33} 
\left(
 \frac{m^2_{\tilde{b},LR}}{m^2_{\tilde{b}_2}-m^2_{\tilde{b}_1}}
\right)^2 \times
\nonumber \\
& & 
\left( B_0(p^2,m^2_{\tilde{b}_1},m^2_{\tilde{b}_1}) + 
       B_0(p^2,m^2_{\tilde{b}_2},m^2_{\tilde{b}_2}) -
     2 B_0(p^2,m^2_{\tilde{b}_1},m^2_{\tilde{b}_2})   
\right) \,,
\label{splittlamprime}
\\
(\delta m_{\tilde{\nu}_{i i^\prime}}^2(p^2))^{(1)}\vert_{\lambda^\prime}
&=& 
 - \frac{3}{4\pi^2} \lambda^\prime_{i33} \lambda^\prime_{i^\prime 33} 
 \, m_b^2 \times
\nonumber \\
& & 
\left( B_0(p^2,m^2_{\tilde{b}_1},m^2_{\tilde{b}_1}) + 
       B_0(p^2,m^2_{\tilde{b}_2},m^2_{\tilde{b}_2}) -
     2 B_0(p^2,m^2_b,m^2_b)   
\right) \,,
\label{splittAprime}
\end{eqnarray}
where the function $B_0(p^2,m_1^2,m^2_2)$ is defined 
as~\cite{Hagiwara:1994pw}:
\begin{equation}
 B_0(p^2,m_1^2,m^2_2) = \int \frac{\overline{d^Dk}}{i\pi^2} 
\frac{1}{(k^2-m_1^2+i\epsilon)[(k+p)^2-m_2^2+i\epsilon]}\,,
\label{bfuncdef}
\end{equation}
where $D=4-2\,\epsilon$ and 
$\overline{d^Dk}=\Gamma(1-\epsilon)(\pi\mu^2)^\epsilon d^Dk$.

\section{Loop functions}
\label{functions}
We list here the functions arising from loop integrations:
\begin{eqnarray}
I(m_1^2,m_2^2,0) & = &                            
  \frac{\ln(m_2^2) -\ln(m_1^2)}  {m_2^2  - m_1^2} \,,
\label{Ifunction}
\\
S(m^2_1, m^2_2)  & = & \frac{2}{(m^2_2-m^2_1)^3}
\left(m^2_2 - m^2_1 + \frac{m^2_2 + m^2_1}{2} 
                      \ln\left(\frac{m^2_1}{m^2_2}\right)
\right) \,,
\label{Sfunction}
\\[1.01ex]
 I^\prime(m^2_1,m^2_2) & = & 
\frac{1-y + y \ln{y}}{\left(1-y\right)^2}\,,
\quad \quad 
 y \equiv \left(\frac{m^2_1}{m^2_2}\right)\,.
\label{Iprimefunction}
\end{eqnarray} 
They are respectively the functions obtained from the fermion-sfermion
loop giving rise to the one-loop neutrino mass contribution; from the
sfermion-sfermion contribution giving rise to the one-loop correction
to sneutrino masses proportional to the $A^\prime$ and $A$ soft
parameters; and from the sneutrino-neutralino loop giving rise to
neutrino mass terms.

The function $J(m_1^2,m_2^2)$ also used in the text is 
\begin{equation}
 J(m_1^2,m_2^2) = 
 \sqrt\frac{m_1^2}{m_2^2} \,I^\prime(m^2_1,m^2_2)\,.
\label{Jfunction}       
\end{equation}

\end{document}